\documentclass[final,5p,times,twocolumn]{elsarticle}

\usepackage{amsmath,amssymb}
\usepackage[version=4]{mhchem}     
\usepackage{graphicx}
\usepackage{siunitx}
\usepackage[colorlinks]{hyperref}
\usepackage{verbatim}
\usepackage{color}

\makeatletter
\def\ps@pprintTitle{%
  \let\@oddhead\@empty
  \let\@evenhead\@empty
  \let\@oddfoot\@empty
  \let\@evenfoot\@empty}
\makeatother

\begin{document}

\begin{frontmatter}

\title{{\bf Chemo-mechanical coupling stabilizes mixed Ag$_{x}$Cu$_{1-x}$GaSe$_2$ solar-cell absorbers:
Insights from Monte-Carlo simulations assisted by \emph{ab initio} informed machine-learning potentials}}

\author{Vasilios Karanikolas\corref{cor1}}
\ead{karanikv@mm.tu-darmstadt.de}
\author{Delwin Perera}
\author{Linus Erhard}
\author{Jochen Rohrer}
\author{Karsten Albe}
\affiliation[inst1]{organization={Institute of Materials Science, Technische Universitat Darmstadt},
  addressline={Otto-Berndt-Stra\ss{}e 3},
  city={Darmstadt}, postcode={D-64287}, country={Germany}}

\cortext[cor1]{Corresponding author.}

\begin{abstract}
Alloying Ag into Cu(In,Ga)Se$_2$ has enabled record solar-cell efficiencies ($\sim23.6\%$), yet their long-term stability remains in question
because {\it ab initio} calculations predict a Ag-Cu miscibility gap near ambient temperature. 
By off-lattice Monte-Carlo simulations using a newly developed machine learning (ML) interatomic potential 
we show that the presence of coherency strain is resolving the controversy between experimental observations and the predicted phase stability. Incorporating elastic energy contributions present in a coherent setup results in complete Ag-Cu miscibility, whereas the expected phase separation occurs in the absence of coherency strains with respect to the end boundary phases, which are mimicked by an incoherent interface with misfit dislocations. The developed ML-MC framework provides a novel approach for resolving discrepancies in 
thermodynamic stability for systems where mechanical and chemical effects compete.
\end{abstract}

\end{frontmatter}


\section{Introduction}
Cu(In,Ga)Se$_2$ (CIGS) thin films exhibit strong absorption in the visible part of the electromagnetic spectrum \citep{Oliveira2022}. This enables lightweight devices 
and integration into building infrastructure and flexible applications \citep{Mufti2020}. The efficiency of CIGS devices has 
remained close to $\sim 20\%$ for roughly a decade, although gains have been reported by alloying Ag to the Cu sublattice 
\citep{Nakamura2019,Keller2024}. Ag alloying lowers the melting temperature \citep{Erslev2011}, can reduce defects \citep{Yang2021}, 
and promotes larger grains \citep{Abou-Ras2016,Valdes2019}, improving the solar-cell device performance \citep{Kim2016,Gu2024}. It also provides 
an additional handle to tune the band gap \citep{Kim2018,Valdes2019} and lattice parameters \citep{Zhang2020}, 
which is attractive for tandem designs \citep{Krause2023}.

Ag is incorporated in the group-I cation sublattice (sublattice-(I)), occupied by Cu in chalcopyrite CIGS. For device applications, 
Ag and Cu should remain homogeneously distributed during deposition, post-deposition processing, and operation \citep{Siebentritt2023}. 
However, \textit{ab initio} calculations combined with a regular-solution model predict a room-temperature miscibility gap 
in Ag$_x$Cu$_{1-x}$GaSe$_2$ \citep{Chen2007,Sopiha2020}, where $x$ is the concentration of Ag to Cu, i.e. $x \rm=[Ag]/([Ag]+[Cu])$.  Accordingly, for ingots synthesized by direct fusion of a stoichiometric mixture of the elements, a miscibility gap of Ag-Cu distribution was reported \citep{Robbins1973,Avon1984,Albornoz2014}.
Also, in thin films obtained from Ag-precursor-assisted processing, significant compositional gradients were observed \citep{Kim2016}.

Elemental co-evaporation is the main fabrication technique for (Ag,Cu)(In,Ga)Se$_2$ thin-films, yields a solid solution of Ag and Cu on sublattice-(I) \citep{Hanket2009,Hanket2010,Chen2014,Boyle2011,Boyle2014}. The absence of gradients in the Ag-Cu distribution in these samples is typically attributed to  slow Ag diffusion \citep{Sopiha2020}, but might also be related to different thermodynamic boundary conditions.

In this study, we revisit the thermodynamics of  Ag$_x$Cu$_{1-x}$GaSe$_2$ by atomistic simulations coupling a machine learning (ML) interatomic potential and a
Monte-Carlo (MC) protocol, which captures the thermomechanics of the system of interest. ML potentials provide {\it ab initio} accuracy at classical potential speeds when implemented in molecular dynamic simulations \citep{Erhard2022,Mortazavi2023,Erhard2024,Jacobs2025}. The resulting 
ML-MC model incorporates lattice parameters and elastic anisotropy beyond Vegard-type interpolations \citep{Baskaran2015}, resulting in 
quantitative assessment of elastic enthalpy contributions.
We construct a DFT database, expand it by on-the-fly active learning, and introduce a MC protocol that combines partial cell 
relaxation, time-stamped force-biased MC (tfMC) \citep{Dereli1992,Timonova2010,Mees2012,Bal2014} and canonical MC. The  relax-tfMC/MC protocol is applied to layered coherent and incoherent initial states.

We find that a key missing ingredient to the thermodynamics of Ag-Cu mixing is the coupled chemo-mechanical equilibrium 
imposed by coherency: local Ag-Cu chemistry interacts with global lattice-parameter constraints, generating coherency strain 
and an elastic contribution to the mixing thermodynamics \citep{Eshelby1957,Larche1973,Larche1982,Larche1984,Fratzl1999,Kamachali2022,Weissmuller2022}. Including the elastic enthalpy in the free energy strongly modifies the phase diagram and yields complete miscibility under coherent constraints. The results explain the discrepancy between the experimental findings and show that stability criteria for chalcopyrites based on lattice-parameters  \citep{Robbins1975} need to be revisited.

The paper is organized as follows: Section~II describes the database construction and active learning; Section~III contains the validation of structural and elastic properties (III\,A), the construction of the composition--temperature phase diagram (III\,B), and relax--tfMC/MC simulations of coherent and incoherent layered structures (III\,C); in Section~IV, semi-analytical and fully numerical models are introduced to calculate the elastic energy contribution to the thermodynamic properties, before we conclude with a summary and outlook in Section~(V).

\begin{figure}
\centering
\includegraphics[width=0.3\textwidth]{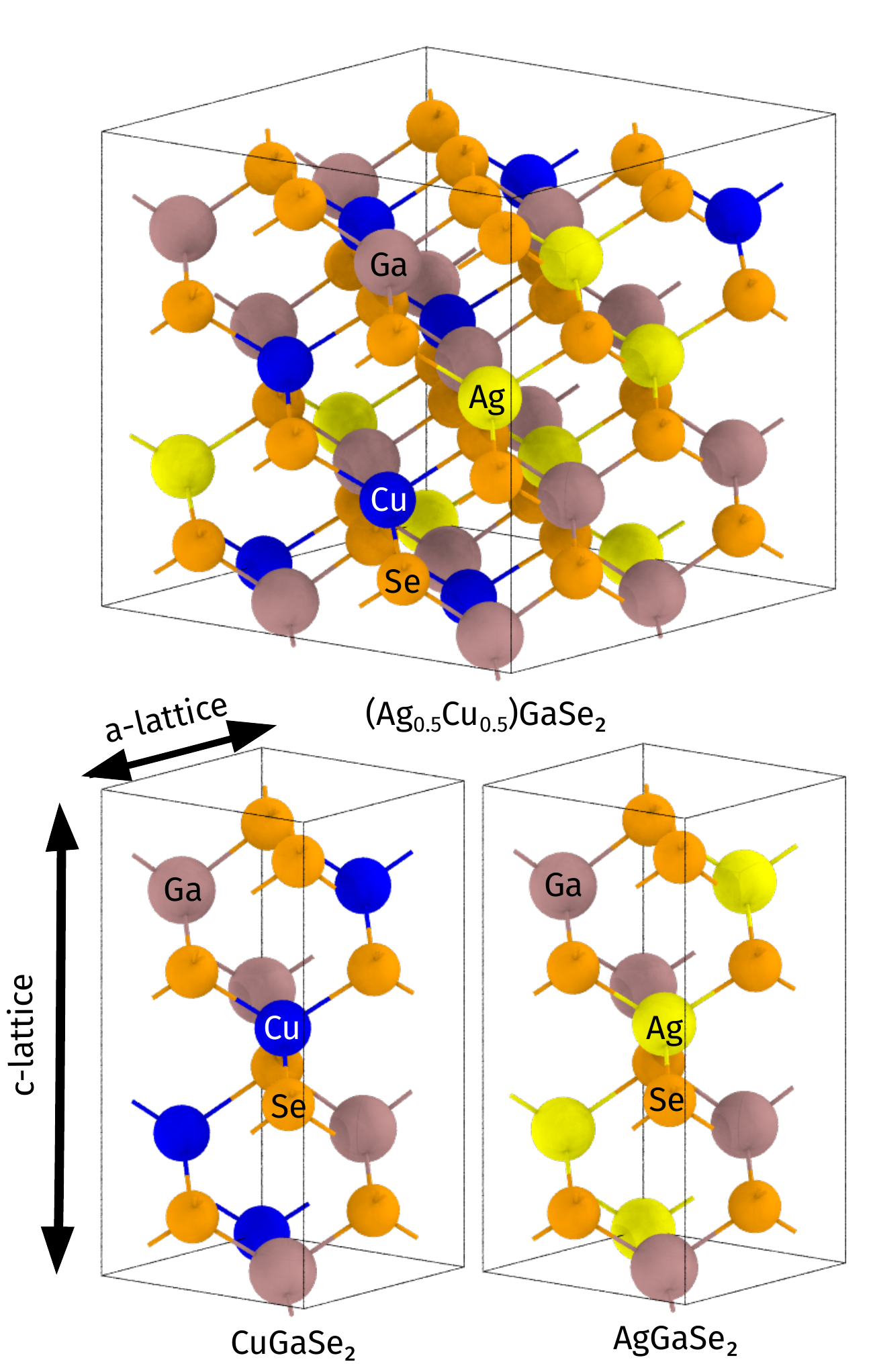}
\caption{ Upper panel: Ag$_{0.5}$Cu$_{0.5}$)GaSe$_2$ structure with Ag-Cu atoms  homogeneously distributed in sublattice-(I). Lower panel: AgGaSe$_2$ and CuGaSe$_2$ structures with $a-$ and $c-$ lattice parameters, which are investigated in Sec.\,III.  \label{fig:Figure01}}
\end{figure}

\section{Methods\label{sec:II}}

\begin{figure*}
\centering
\includegraphics[width=0.8\textwidth]{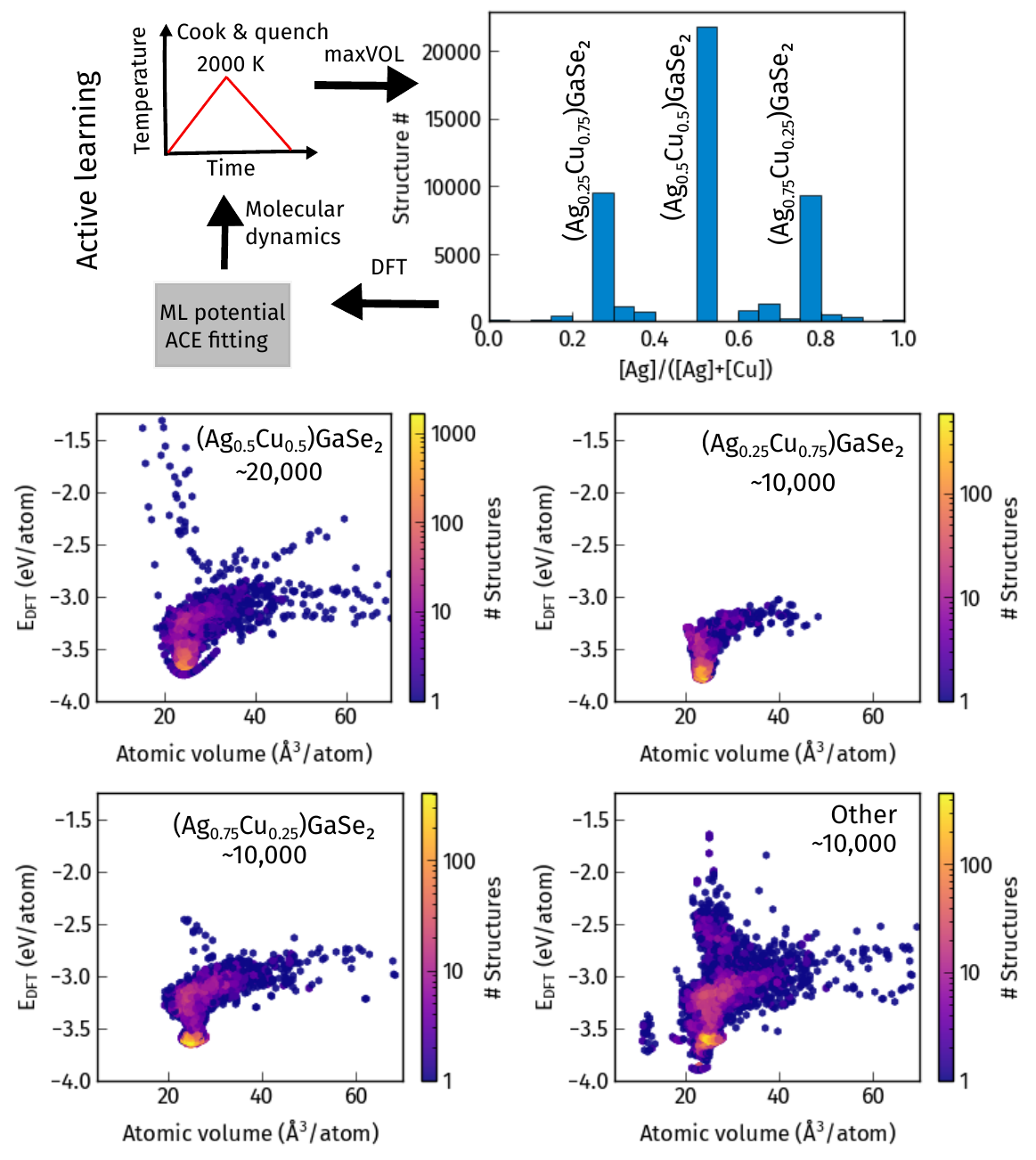}
\caption{ We present the workflow used to develop the machine-learning (ML) potential. Molecular-dynamics (MD) simulations are performed following a cook-and-quench protocol, considering multiple Ag-Cu concentrations, defined by the ratio $[{\rm Ag}]/([{\rm Ag}]+[{\rm Cu}])$. Several active-learning rounds are applied to iteratively explore the parameter space of Ag$_x$Cu$_{1-x}$GaSe$_2$ and expand the data set with structures that have extrapolation grades between 5 and 20 using the maxVOL selection scheme, yielding a faithful ML potential across the targeted phases. From the distribution of the number of generated structures as a function of composition, we observe that the main focus is on $x=0.25$, $0.5$ and $0.75$ phases. Finally, we show atomic volume versus atomic energy scatter plots to illustrate the spread of the data set in the parameter space over successive active-learning iterations.\label{fig:Figure02}}
\end{figure*}

As a first step, the machine learning interatomic potential is developed to predict structural, mechanical, and thermodynamic properties. 
Thereafter, we study the delicate interplay between the local chemical composition and the coherency lattice 
parameters on the phase diagram of Ag$_x$Cu$_{1-x}$GaSe$_2$ mixtures.

The ML potential is fitted using the ACE formalism for the quaternary of (Ag$_x$Cu$_{1-x}$)GaSe$_2$  
\citep{Drautz2019}. The ACE formalism is chosen because it offers a good compromise between accuracy, speed,  and ease of 
implementation  \citep{Lysogorskiy2021,Qamar2023,Leimeroth2025}. The ML potential predicts the 
atomic energy and forces acting upon an atom by its local environment in a given configuration.
The full database (DB) includes about 1 million structurally and chemically different atomic environments. The DB is built through active learning
cycles  using the maxVOL selection scheme \citep{Podryabinkin2017,Novikov2021,Lysogorskiy2023} as implemented in 
the PACEMAKER code \citep{Lysogorskiy2021}. Figure~\ref{fig:Figure02} schematically shows the workflow and visualizes the data in energy-volume space.
After DFT calculations were performed, the first ML potential version was fitted. Then, the active learning process is applied, 
where cook-and-quench molecular dynamics (MD) simulations are  performed. Structures with 
extrapolation grades between 5 and 20 are collected and eventually used to train the next version 
of the ML potential, after performing DFT calculations on the selected arrangements. 

Since the Ag$_x$Cu$_{1-x}$GaSe$_2$ is in the tetragonal phase (see Figure \ref{fig:Figure01}), we focused on this phase in the first active 
learning cycles; 21,714 structures have the composition (Ag$_{0.5}$Cu$_{0.5}$)GaSe$_2$, 9,474 structures have 
the composition (Ag$_{0.25}$Cu$_{0.75}$)GaSe$_2$ and 9,143 structures have the composition (Ag$_{0.75}$Cu$_{0.25}$)GaSe$_2$ as shown in Figure \ref{fig:Figure02}. 
Moreover, we considered $164$ 
input structures of different Ag to Cu ratio, $x=[0,1]$, with 12 tetragonal, 64 monoclinic, 81 triclinic and 
7 orthorhombic phases. For each of these structures, we apply cell 
deformations and atom displacements at random \citep{Bochkarev2022}. 
The rest of the DB includes all the other phases and a subset that has been used to test the 
predictions of the ML potential. In particular, $\sim30$ structures that started from the pure BCC Cu 
phase and $\sim100$ structures were extracted by testing the energy barrier for Ag and Cu vacancies
using the NEB method. 
Molecular dynamics simulations were performed to test Ag and Cu diffusion via the vacancy mechanism. For this, 
structures  with $2398$ atoms were tested at $1200\,$ and $1300\,$K. The amorphous matrix embedding method, 
from Ref.\,\citep{Erhard2024}, is applied to cut small-scale motifs ($<100$ atoms) around atoms with high extrapolation grade 
to perform DFT calculations, leading to another $\sim 1000$ structures in the DB.

For all structures in the DB, DFT calculations are performed using VASP \citep{Kresse1994,Kresse1996} with
the Perdew-Burke-Ernzerhof (PBE) \citep{Perdew1996} exchange correlation
functional and projector augmented wave pseudopotentials \citep{Kresse1999}. A
plane wave energy cutoff of $500\,$eV and k-points spacing of $0.1725\, \text{\AA}^{-1}$ is used
to ensure the convergence of the mixing energy over the Ag to Cu ratio in the $2\times 2\times 1$ Ag$_x$Cu$_{1-x}$GaSe$_2$ tetragonal supercell. 

To apply the cook-and-quench protocol in the active learning we perform molecular dynamic
simulations using the LAMMPS code \citep{Thompson2022}. We perform isothermal-isobaric (NPT) MD simulations using Nose-Hoover thermostat, with relaxation time of $0.1\,$ps. For each input structure we vary the temperature from $300\,$K to
$2000\,$K, keeping the temperature fixed at $2000\,$K for $0.2\,$ps and then
back to $300\,$K, within a time span of $0.4\,$ns, with a time step of $1\,$fs.

For Monte Carlo (MC) simulations we also employ the LAMMPS code. We focus on the Ag and Cu distribution to the 
sublattice-(I) to find stable configurations. We use the atom/swap canonical MC fix to explore the configurational space 
of the Ag and Cu distribution \citep{Sadigh2012}; Ag and Cu atoms are randomly swapped and the Metropolis criterion dictates the swap 
probability, that depends on the scaling temperature. Finally, we utilize the ML 
potential and combine time-stamped forced-bias MC (tfMC) \citep{Mees2012,Bal2014,Neyts2013} with canonical MC for the Ag and Cu swaps and partial 
relaxation of the atomic positions and considered supercell; from now on this method is named relax-tfMC/MC protocol.

Validation graphs for the ML potential can be found in the Supplemental Materials section \citep{SupplementalMaterials}. Here, we only briefly 
report that the ML potential exhibits mean absolute errors (MAEs) of 3.1 meV/atom and 60.9 ~meV/\AA~ 
and root mean square errors of $5.7\,$meV/atom and $98.1\,$meV/$\text{\AA}$  for energies and forces, 
respectively (see Figure S1 for more details). For the boundary phases AgGaSe$_2$ and CuGaSe$_2$ as well as for the 
ordered (Ag,Cu)GaSe$_2$ phase, the energy volume curves present MAE values $<1\,$meV/atom from the corresponding 
DFT data. Furthermore, the derived bulk moduli agree within less than $5$ GPa with DFT values, see Figure S2 and Tab. S1 \citep{Grimsditch1975,Neumann1983,Neumann1986,Fernandez1990}.

\section{Results \label{sec:III}}

\subsection{Validation of structural properties}

The first validation step of the ML potential comprised the compositional dependence of structural parameters of Ag$_x$Cu$_{1-x}$GaSe$_2$. Using the ML potential, 
we calculate the lattice parameters for  $16\times 16 \times 8$ Ag$_x$Cu$_{1-x}$GaSe$_{2}$ supercells, which
consist of $32,768$ atoms. Ag and Cu atoms occupy  sublattice-(I) and
are randomly distributed to mimic a solid solution. 
For the DFT calculations we use a $2\times 2\times 1$ supercell also with a random distribution of Ag and Cu atoms. In order to test the sensitivity of our results on finite-size effects, in addition, we calculated the lattice parameters with the ML potential using the same $2\times 2\times 1$ supercells.

\begin{figure}
\centering
\includegraphics[width=0.45\textwidth]{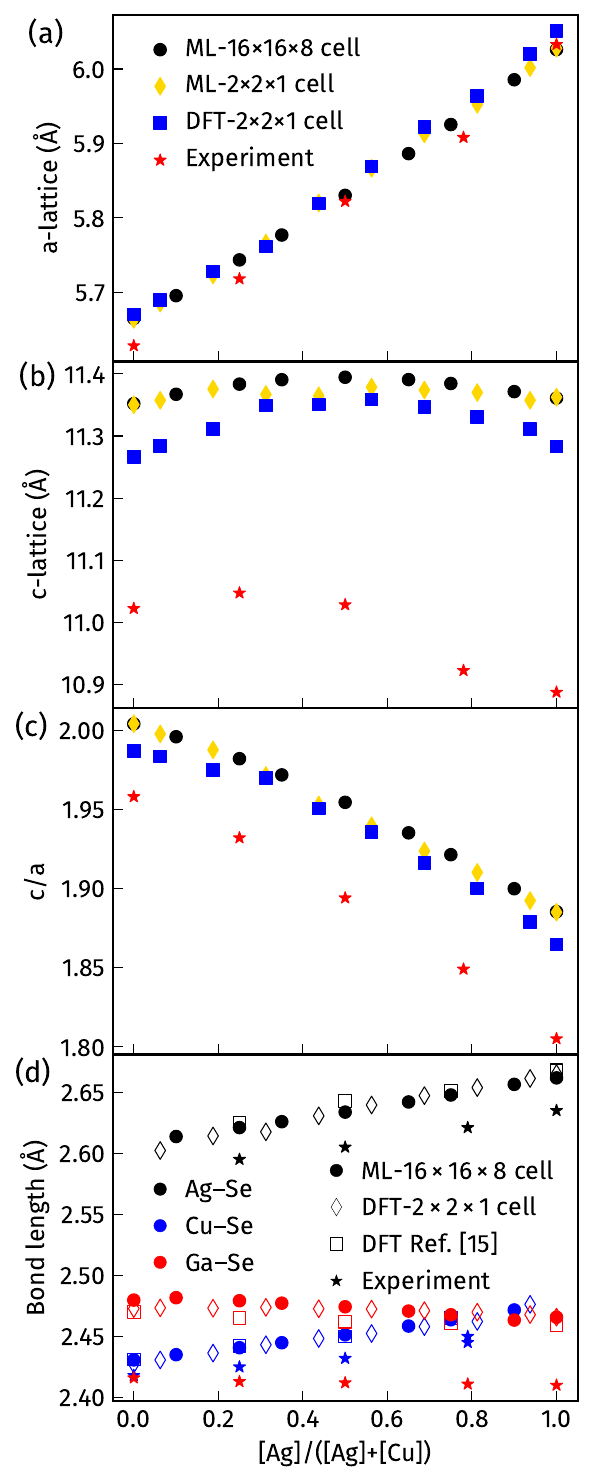}

\caption{Calculated (a) $a$ lattice parameter, (b) $c$ lattice parameter, (c) the $c/a$ values and (d) anion-cation bond lengths for the different Ag to Cu concentrations. The predicted values by the ML potential are extracted 
by relaxing $16\times 16 \times 8$ solid solution structures. Our DFT calculations in (a-d) are
obtained using a $2\times 2\times 1$ supercell with random Ag-Cu distribution. DFT data shown in panel (d), with square symbols, are taken from Ref.\,\citep{Chen2007}. The experimental results for the (a-c) lattice parameters are from Ref.~\citep{Boyle2014} and for the (d) bond lengths are from Ref.\,\citep{Falk2023}.
\label{fig:Figure03}}
\end{figure}

Figure \ref{fig:Figure03}(a) shows that the static $a$-lattice parameter increases with increasing Ag content and 
follows the data obtained from DFT calculations. There is also good agreement with the experimental values from Ref.~\citep{Boyle2014}, which are, however, taken at room temperature. 
Also, the non-linear dependence of the $c$-lattice parameter on the Ag content is captured by the 
ML potential and DFT training data, even though the quantitative agreement is not as good as for the a-lattice parameter (see Figure \ref{fig:Figure03}(b,c) ).
Comparing the results for the $2\times 2\times 1$ and $16\times 16\times 8$ supercells obtained with the ML potential reveals a significant size dependence of the result. Also, the data deviate from the DFT reference, which, however, were not part of the training data set. Given that the relative error of the calculated c-axis is on the order of the observed change with composition, one can conclude the the compositional dependence of the cell parameter $c$ is insignificant.

Finally, the $c/a$ ratio calculated using the ML potential
nicely reproduces the increasing tetragonal distortion with increasing Ag content.
The $c/a$ value obtained from the ML potential
closely follows the experimental data and the valued from DFT calculations.
The offset of the theoretical data can be attributed to the general underbinding of the 
PBE functional.

Ag-Cu alloying also leads to modifications of the bond lengths,
which are accompanied by anion displacements \citep{Garbato1987}. 
Figure~\ref{fig:Figure03}\,(d) shows the Ag-Se, Cu-Se and Ga-Se bond lengths for Ag$_x$Cu$_{1-x}$GaSe$_2$ bulk structures as a function of the Ag content.
Ag and Cu are randomly distributed on the  sublattice-(I). We observe that the Ga-Se bond
length exhibits  only very small variations with increasing Ag to Cu ratio,
compared to Ag-Se and Cu-Se. This is expected because the Ga sublattice remains unchanged 
and Ga retains tetrahedral coordination by Se, so its local environment is affected only 
indirectly by the alloy composition. For both, Ag-Se
and Cu-Se interactions, we observe increasing bond lengths with
increasing Ag content. This trend is consistent with the larger effective size of Ag compared to Cu and 
the generally more compliant sublattice-(I)--Se bonds compared to Ga-Se.
The bond lengths predicted for Ag$_x$Cu$_{1-x}$GaSe$_{2}$
structures using the ML potential are in accordance with {\it ab initio} calculations
(open squares) reported in Figure\,\ref{fig:Figure03}\,(d) from Ref.~\citep{Chen2007} and our DFT calculations.
Experimental bond lengths  presented by the star symbols were estimated from XRD data \citep{Falk2023}. 

The ML potential, ab initio calculations, and
experimental data show similar trends in bond lengths as a function of Ag to Cu ratio.
However, the bond lengths predicted by the ML potential and obtained from DFT are generally overestimated compared to the experimental results.
In addition to the presence of disorder, strain, and defects in fabricated ACGS thin films, systematic differences 
can also arise from the approximations inherent to the chosen DFT functional and finite-temperature effects in experiment.

\subsection{Validation of  elastic properties}

\begin{figure}
\centering
\includegraphics[width=0.45\textwidth]{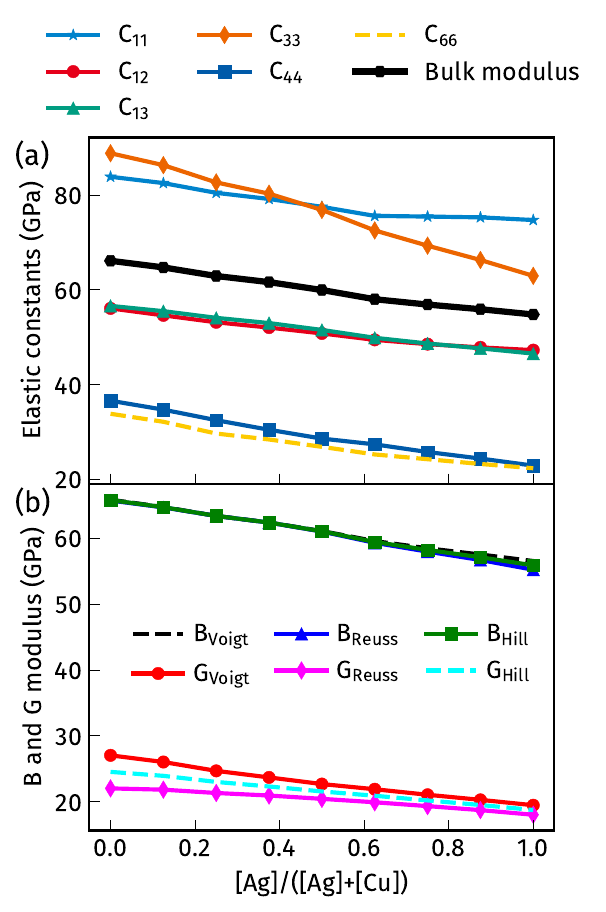}

\caption{(a) The components of the stiffness tensor $C_{ij}$ and the bulk modulus $B$ are calculated using the ML potential, for varying Ag to Cu ratios. (b) The elastic constants $C_{ij}$  are used to calculate the effective moduli for a polycrystalline sample using the Voigt, Reuss and Hill measures. All calculation are done with $4\times 4\times 2$ supercells and fully relaxed  structures with different Ag to Cu ratios. \label{fig:Figure04} }
\end{figure}

In order to capture thermomechanical effects in Ag$_x$Cu$_{1-x}$GaSe$_2$ the stiffness tensor is an important quantity.
Up to now, in the literature, there is only data available for the  AgGaSe$_2$ and CuGaSe$_2$ 
boundary phases. As an example, for CuGaSe$_2$  and AgGaSe$_2$ calculated values of $C_{11}$=87.3 and $C_{11}$=83.6 have been reported \citep{Verma2012}, respectively, while another source is giving values of  $C_{11}$=105.4\,GPa and $C_{11}$=90.4\,GPa \citep{Kushwaha2019}.
In order to obtain consistent data, we calculated the components of the stiffness tensor as function of composition for a fully relaxed $4\times 4\times 2$ supercell,
where the Ag-Cu atoms were randomly distributed. 
Symmetric lattice deformations of $\delta=\pm 0.01\,\text{\AA}$ were applied
to the $a$ and $c$ lattice parameters, before we relaxed the internal atomic positions. 
We then compute the corresponding stress components, $\sigma{\pm}$, and evaluate
$C_{ij}=(\sigma_{ij,+}-\sigma_{ij,-})/(2|\delta|)$, using the central finite-difference estimate 
of $\partial\sigma_{ij}/\partial\varepsilon$.   

In Figure \ref{fig:Figure04}\,(a) the calculated moduli are shown  as a function 
of Ag content, $x=$[Ag]/([Ag]+[Cu]), together with the isotropic bulk modulus, $B=1/9\left[2(C_{11}+C_{12})+4C_{13}+C_{33}\right]$. In
Figure \ref{fig:Figure04}(b) effective moduli representative for polycrystalline samples are also shown. The Voigt estimate corresponds to uniform strain across grains, whereas the Reuss estimate corresponds to uniform stress. The Voigt values provide an upper bound, the Reuss values a lower bound, and the Hill average is the arithmetic mean of the two.

All components $C_{ij}$ 
decrease with increasing Ag content, indicating elastic softening with increasing Ag-content.
This can be qualitatively understood:
since Cu-Se bonds are shorter than Ag-Se bonds, this is implying a larger restoring force for Cu-Se bond stretching and hence a larger stiffness. Therefore, increasing the Cu fraction is expected to increase the curvature of the potential energy surface with respect to atomic displacements, resulting in larger $C_{ij}$ values (Figure \ref{fig:Figure04}\,(a)). Moreover, Ag has a larger ionic/atomic size than Cu, so the lattice expands as Ag content increases. Larger interatomic distances reduce orbital overlap and weaken bonding, which lowers the stiffness; accordingly, the bulk modulus decreases and the lattice becomes more compressible.

\subsection{Bulk thermodynamic properties}
\begin{figure*}[h]
\centering
\includegraphics[width=1.0\textwidth]{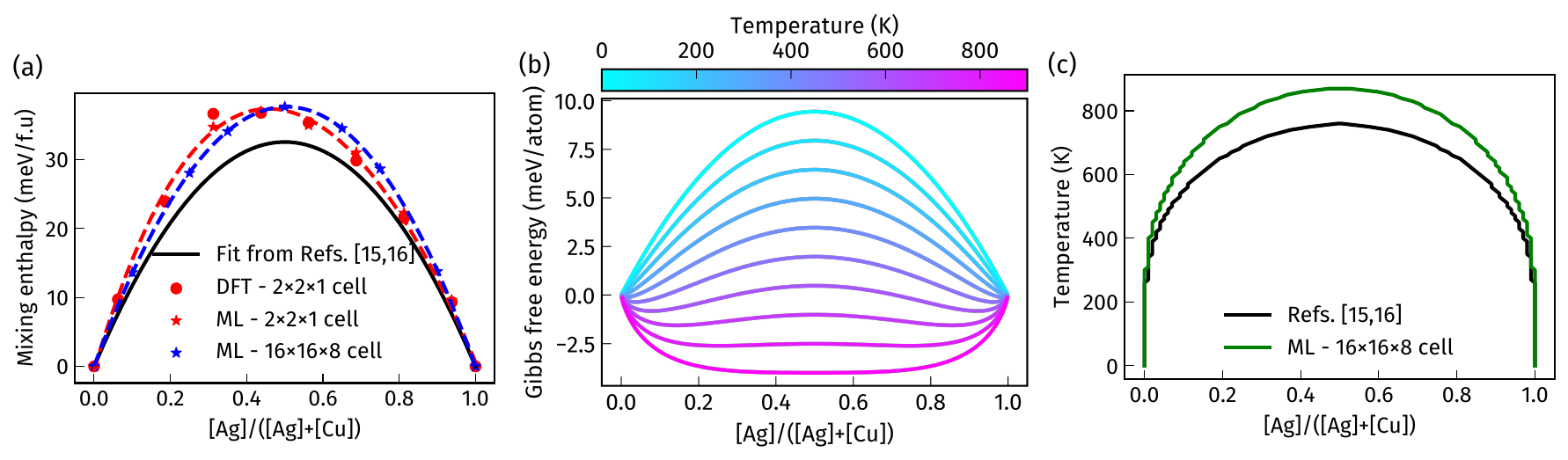}

\caption{(a) Mixing enthalpy per formula unit for varying Ag to Cu ratios; the
circular points give the DFT values and the star-shaped points give the predicted
values from the ML potential. The parabolic fits are
superimposed. The black line gives the analytical fit from
Refs.~\citep{Chen2007,Sopiha2020}. (b) The Gibbs free energy $G_{\text{mix}}$ per atom, varying Ag to
Cu ratios, predicted from the ML potential, using
the fitting from Eq.~\ref{eq:03} for various temperatures $T$. (c) Pseudo-binary phase diagram obtained from a common tangent construction using the
ML potential to calculate the mixing energies in the  $16\times 16\times 8$ Ag$_x$Cu$_{1-x}$GaSe$_2$ supercells
and the analytical fit from Refs.\,\citep{Chen2007,Sopiha2020} \label{fig:Figure05}}
\end{figure*}

The efficiency of the ACGS thin-film solar cell devices is highly affected by Ag and Cu configuration 
in sublattice-(I), where fabricating and maintaining the thin film in solid-solution configuration is the goal. 
Thus, we turn now our attention to the thermodynamic properties 
of the Ag$_x$Cu$_{1-x}$GaSe$_2$ bulk structure and calculate the mixing enthalpy $H$. The Gibbs free energy $G_{\text{mix}}$ 
within the regular solution model is defined as,
\begin{equation}
\Delta G_{\text{mix}}=\Delta  H_{\text{mix}}-T \Delta S_{\text{mix}},
\end{equation}
where $ \Delta  H_{\text{mix}}$ is the mixing enthalpy for the studied system,
$S_{\text{mix}}$ is the ideal configurational entropy of mixing, and $T$ is the
temperature. The Ag and Cu atoms occupy the sublattice-(I), thus the mixing enthalpy is calculated for
a Ag concentration $x$ as
\begin{equation}
 \Delta H_{\text{mix}}=H_{\text{Ag}_{x}\text{Cu}_{1-x}\text{GaSe}_{2}}-(1-x)H_{\text{CuGaSe}_{2}}-xH_{\text{AgGaSe}_{2}},
\label{eq:02}
\end{equation}
where $H_{\text{Ag}_{x}\text{Cu}_{1-x}\text{GaSe}_{2}}$ is the enthalpy of the random solid solution; 
$H_{\text{CuGaSe}_{2}}$ and $H_{\text{AgGaSe}_{2}}$ are the reference  enthalpies of the boundary phases.
Then, $ \Delta  H_{\text{mix}}$ can be fitted by the expression,
\begin{equation}
\Delta H_{\text{mix}}=\Omega x(1-x),\label{eq:03}
\end{equation}
where $x=[\text{Ag}]/([\text{Ag}]+[\text{Cu}])$ is the Ag to Cu ratio
and $\Omega$ is the interaction parameter. 

The ideal entropy of mixing for the binary system is given by the following analytical expression:
\begin{equation}
    \Delta S_{\text{mix}}=k_{B}\left[x\ln(x)+(1-x)\ln(1-x)\right].\label{eq:04}
\end{equation} 
To assess the thermodynamic stability, the binodal lines are calculated from a
common tangent construction of the resulting Gibbs free energy of mixing.

Using Eq.~\ref{eq:02} we calculate the mixing enthalpy values and
subsequently we apply the parabolic fitting of Eq.~\ref{eq:03} using
values from DFT calculation and the values predicted by the ML potential.
The DFT calculations are performed in the $2\times 2\times 1$ supercell and for the same structures
we use the ML potential to calculate the mixing enthalpy at zero pressure.
The ML potential is also used in the $16\times 16\times 8$ supercell.
The mixing enthalpy values and parabolic fits are presented
in Figure~\ref{fig:Figure05}(a). The
maximum mixing enthalpy  per formula unit is around 32 meV at $x_{Ag}=0.5$, which 
corresponds to a critical spinodal temperature $>$ 800 K, indicating the tendency for phase decomposition at room temperature.
The data obtained from the ML potential and the ab initio data are close, even though we are operating on an energy scale of only a few meV/f.u. Also, there is a good agreement with the DFT calculations from Refs.\,\citep{Chen2007,Sopiha2020}, where slightly different parameters and smaller cells were used.

Based on the parabolic fitting, Eq.~\ref{eq:03}, for the
mixing enthalpy $ \Delta H_{\text{mix}}$ and the mixing entropy $\Delta S_{\text{mix}}$, Eq.~\ref{eq:04},
we calculate the Gibbs free energy of mixing $ \Delta G_{\text{mix}}$
In Figure \ref{fig:Figure05} (b) the result is shown
for various temperatures as function of the Ag content. 

In Figure~\ref{fig:Figure05}\,(c) we present the resulting pseudo-binary phase diagram
of Ag$_{x}$Cu$_{1-x}$GaSe$_{2}$, where the solid lines separate the two-phase
region from the solid solution. We see that the {\em incoherent} miscibility gap extends over a
wide compositional range. At $300$\,K the solubility of Ag in Cu is only a small percentage 
($<1\%$). The result confirms earlier studies \citep{Chen2007,Sopiha2020}, which predict a slightly lower critical temperature.
It is important to note at this point that the thermodynamic analysis up to this point is ignoring the impact of interfaces, which can cause coherency strains. This is the subject of the following sections.

\subsection{Interplay of mechanics and local chemical composition}

The predicted phase separation of the Ag and Cu atoms at room temperature is clearly in conflict with the 
long-term performance stability of ACGS solar cell devices \citep{Hanket2009,Hanket2010}. 
This performance stability necessitates the persistence of single-phase mixed ACGS. 
In this section, we demonstrate that this persistence results from a chemical-mechanical coupling. We investigate the Ag and Cu
compositional distribution in sublattice-(I) utilizing the ML potential in large-scale MC simulations. We use the relax-tfMC/MC
protocol which is presented in the Supplementary Material \citep{SupplementalMaterials}. The Warren-Cowley (WC) short range order parameter is introduced to understand 
the mixing of Ag and Cu in sublattice-(I); for the fully mixed Ag-Cu distributions the WC parameter has values 
close to $0$, for the phase-separated structures the WC values tend to $1$. All visualizations and analysis of the WC order parameter structural features, OVITO PRO is used \citep{Stukowski2010}.  In the following we apply the relax-tfMC/MC protocol 
to a coherent and an incoherent interface structures to find equilibrated Ag-Cu distributions. The coherent interface has the 
global lattice parameters of the homogeneous (Ag$_{0.5}$Cu$_{0.5}$)GaSe$_2$; the incoherent interface is engineered so as to have two areas with two global lattice parameters that match the pure phase AgGaSe$_2$ and CuGaSe$_2$.

\begin{figure*} 
\centering
\includegraphics[width=0.95\textwidth]{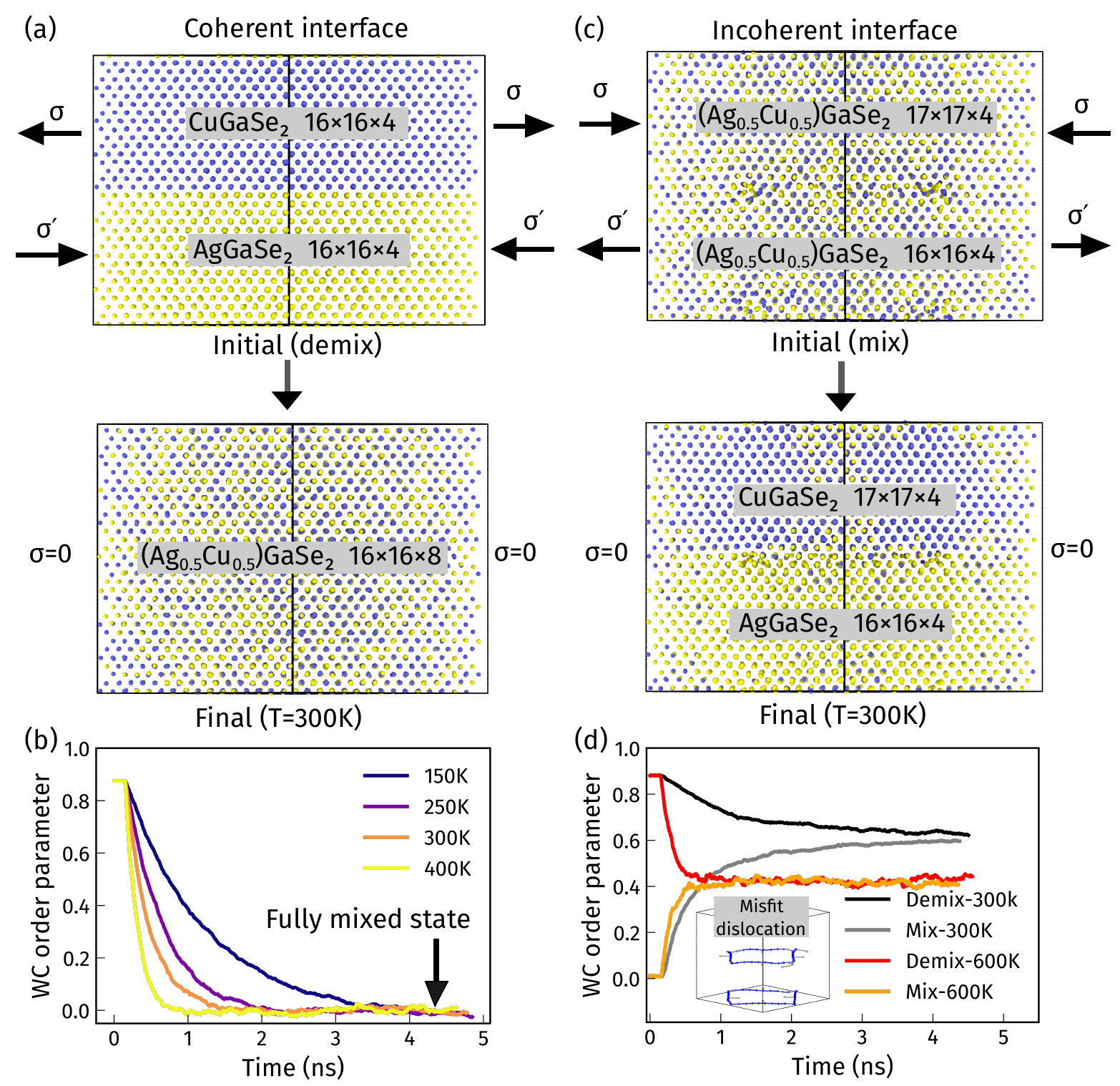}

\caption{ Equilibration of a coherent and incoherent  layered interfaces using the relax-tfMC/MC protocol. In both structures a strain is induced due to the mismatch between the local Ag-Cu composition and the global coherency lattice parameters. (a) Initial and final Ag-Cu configurations for a coherent layered heterostructure constructed from $16\times 16\times 4$ AgGaSe$_2$ and 16$\times$16$\times$4 CuGaSe$_2$ layered supercells. The input structure is initially phase separated (Ag- and Cu-rich layers) and has homogeneous (Ag$_{0.5}$Cu$_{0.5}$)GaSe$_2$ lattice parameters. The relax-tfMC/MC protocol is applied and  Ag-Cu become fully mixed, driven by coherency stress. (b) Time evolution of the Warren--Cowley short-range order parameter for the initial configuration in (a) at $T=150$, $250$, $300$, and $400$~K presents that the system is in a fully mixed phase.
(c) Initial and final Ag-Cu configurations for an incoherent-interface geometry constructed from
$16\times 16\times 4$ AgGaSe$_2$ and $17\times 17\times 4$ 
CuGaSe$_2$ supercells. The input structure is in fully mixed phase and by construction has the two pure phase lattice parameters. The relax-tfMC/MC protocol is applied and the system phase separates, consistent with the reduced coherency constraint in this geometry.
(d) Time evolution of the Warren--Cowley order parameter for the incoherent-interface case, starting from two initial configurations (fully mixed and phase separated) built from the end-point lattice parameters, at $T=300$ and $600$~K.; we observe the tendency of the system to be phase separated. \label{fig:Figure06}}

\end{figure*}

\subsubsection{Coherent interface}

We start with a coherent layered model structure where the Ag and Cu atoms are initially phase separated and the global lattice parameters matches that of a fully mixed (Ag$_{0.5}$Cu$_{0.5}$)GaSe$_2$ solid solution. The interface as shown in Figure \ref{fig:Figure06}(a) is coherent. The in-plane lattice parameter remains continuous across the interface, while the chemical composition changes abruptly.
Consequently, in the coherent layered configuration the lower AgGaSe$_2$ 
layer is initially compressed relative to its equilibrium state, whereas the upper CuGaSe$_2$ region is elastically stretched. This coherency strain generates an internal stress in the initial configuration as shown in Figure \ref{fig:Figure06}(a).

When we apply the relax-tfMC/MC protocol at room temperature ($300\,$K), the system evolves from the demixed configuration toward a 
fully mixed stress free phase. This behavior contrasts 
with the incoherent mixing enthalpy shown in Figure \ref{fig:Figure05}, where the elastic energy associated with the coherency strain is not considered.
Figure \ref{fig:Figure06}(b) shows that, full mixing is observed at all temperatures from $150$ to $400$ K. In all cases, the system evolves from the initially phase-separated configuration to a fully mixed Ag-Cu distribution on sublattice-(I).

\subsubsection{Incoherent interface}

Now we consider a structure that is composed of two layers separated by an incoherent interface. 
The layered supercell is constructed from a sandwich of $16\times 16\times 4$ and $17\times 17\times 4$ cells.
The structure is designed to approximately satisfy the coincidence condition $16 a_{\mathrm{AgGaSe_2}}\approx17 a_{\mathrm{CuGaSe_2}}$.
 Because the two sides contain different numbers of 
in-plane unit-cell repeats, the lattice mismatch is accommodated by  misfit dislocations, visible in the inset of Figure \ref{fig:Figure06}(d), which relax the 
transition between the incoherent layers.  The interface is incoherent (or semi-coherent) because  in-plane lattice constants change from the lower to the upper layer.
If we start with a solid solution in both layers as shown in 
Figure \ref{fig:Figure03}(a), there is a stress gradient between the upper and lower layer: the local chemical composition does not match with the global lattice parameters, which generates internal stress in the initial 
configuration of Figure \ref{fig:Figure06}(c). To reduce the excess elastic energy, the system is driven towards demixing, i.e., toward locally 
Ag-rich and Cu-rich regions that better match the respective end-member lattice constants, as seen in Figure \ref{fig:Figure03}(a). 
When we apply the relax-tfMC/MC protocol at room temperature ($300\,$K), the structure evolves to a demixed state,
verifying that full demixing occurs in the absence of coherency strains.

Because of the mismatch of lattice planes, there is initial stress that drives the system out of the miscibility gap from Figure~\ref{fig:Figure05}(c). Again, we apply the relax-tfMC/MC protocol at $300\,$K and $600\,$K 
and compute the WC short-range order parameter over time. Now we consider two initial configurations: one fully mixed and one phase separated, both constructed with 
the same $16\times 16\times 4$ AgGaSe$_2$ and $17\times 17\times 4$ CuGaSe$_2$ geometry (i.e., using the end-member in-plane lattice parameters on the 
respective sides). Figure \ref{fig:Figure06}(d) shows that starting from a fully mixed Ag-Cu distribution the system evolves toward a phase-separated
configuration at both temperatures, whereas starting from a phase-separated configuration the structure remains phase separated. We find that reducing the stress requires the local composition to move from fully mixed Ag-Cu configuration, towards local chemical compositions that are compatible with lattice constants of the boundary phases.

\section{Analytical modeling for the elastic energy and discussion}

The relax-tfMC/MC protocol with the ML potential unveils a rich underlying physical process, where the phase stability is controlled by the induced strain: 
local composition fluctuations are constrained to adopt the macroscopic (coherent) lattice parameters. The associated local elastic energy must be included when investigating the thermodynamic properties.
Approaches based on fully relaxed energies effectively omit this coherency constraint, which can lead to misleading conclusions. Our goal is to calculate the induced coherency stress, the developed elastic energy and the influence they have on the Gibbs free energy.

\begin{figure*}
\centering
\includegraphics[width=0.75\textwidth]{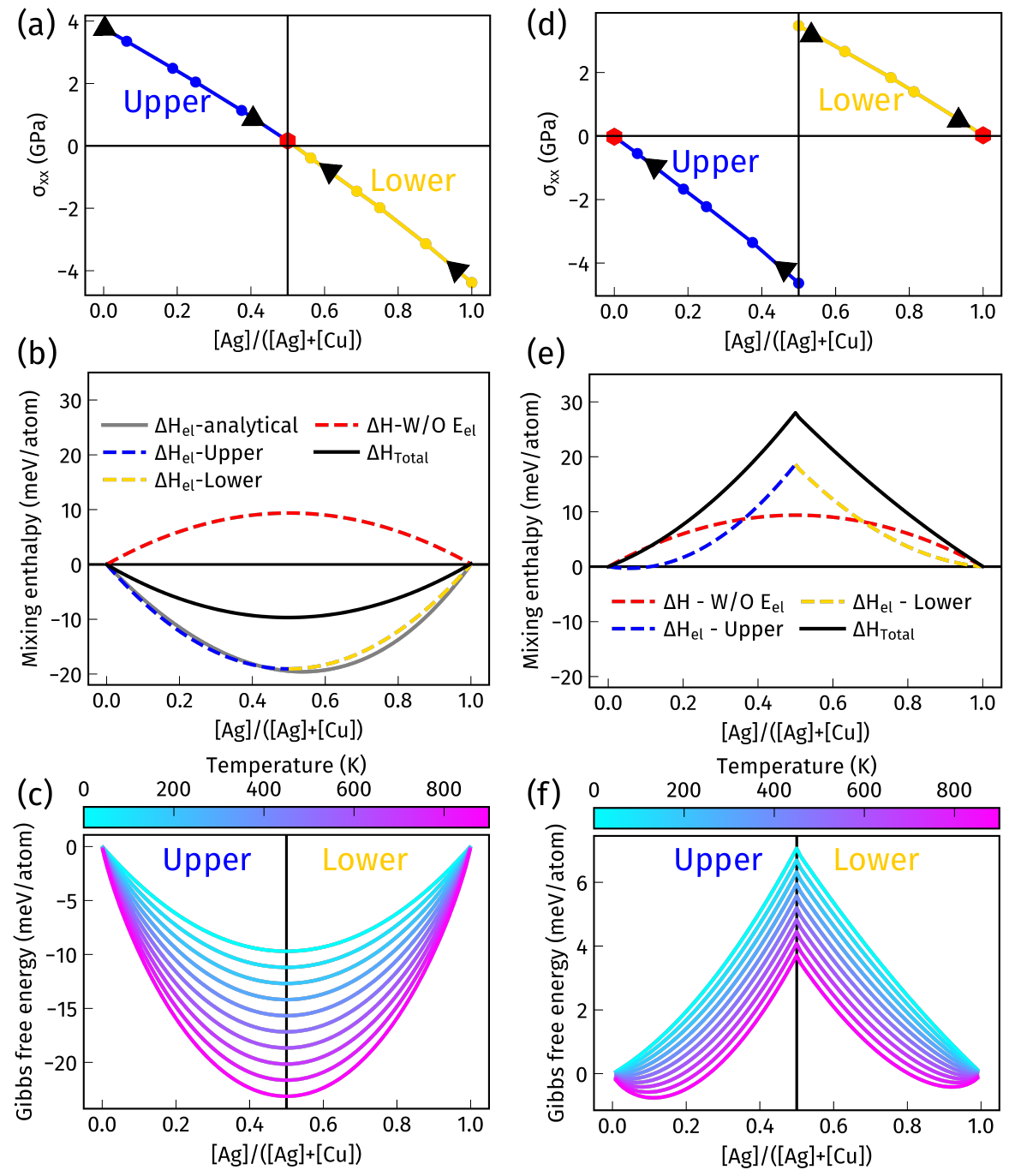}
\caption{  Panels (a-c) use as a reference the (Ag$_{0.5}$Cu$_{0.5}$GaSe$_2$) coherent lattice parameters, whereas panels (d-f) use the end-point references corresponding to AgGaSe$_2$ and CuGaSe$_2$. Panels (a,d) show the induced coherency strain, panels (b,e) show \(\Delta H_{\rm mix}\), \(H_{\rm el}\), and $H_{\rm total}$, and panels (c,f) show the Gibbs free energy of mixing at different temperatures. All quantities are reported as a function of composition (Ag to Cu ratio).
\label{fig:Figure07}} 
\end{figure*}

In the following, we consider a \(4\times 4\times 2\) Ag$_x$Cu$_{1-x}$GaSe$_2$ supercell and impose fixed coherent reference lattice parameters ($a_0$,$c_0$). In Figure~\ref{fig:Figure07}(a) we calculate the in-plane stress $\sigma_{xx}$ induced by coherency strain when ($a_0$,$c_0$) is chosen as the relaxed lattice parameters of (Ag$_{0.5}$Cu$_{0.5}$)GaSe$_2$, while varying the Ag to Cu ratio. As expected, the stress is minimized (and vanishes within numerical precision) at $0.5$; compositions away from $0.5$ develop tensile or compressive stress due to the lattice mismatch with the reference coherent lattice parameters. This choice of global lattice parameters corresponds to the coherent interface scenario in Figure~\ref{fig:Figure06}(a), where reducing the coherency stress drives the system toward a local homogeneous Ag-Cu distribution that matches the imposed lattice parameters for the $0.5$ composition. The arrows in Figure~\ref{fig:Figure07}(a) indicate the direction of the local compositional change to reduce $\sigma_{xx}$, which is the final configuration in Figure~\ref{fig:Figure06}(a) after applying the relax-tfMC/MC protocol. The 'upper' and 'lower' descriptions in Figure \ref{fig:Figure07} are used to define the geometrical orientation from Figures \ref{fig:Figure06}(a,c).

In Figure~\ref{fig:Figure07}(d) we calculate the \(\sigma_{xx}\) as a function of composition using two different coherent references, namely the lattice parameters of the pure phases AgGaSe$_2$ and CuGaSe$_2$. In each case, the stress is zero near the matched boundary phase and its magnitude increases as the composition deviates from the endpoint, inducing the growing coherency strain. The resulting upper and lower branches in Figure~\ref{fig:Figure07}(d) are connected to the incoherent interface scenario of Figure~\ref{fig:Figure06}(c): to reduce the induced strain, the local composition shifts toward the reference lattice parameters in each region, in the upper part becomes Cu-rich and in the lower part becomes Ag-rich). The arrows in Figure~\ref{fig:Figure07}(d) indicate the direction of local compositional change that reduces the coherency strain, consistent with the final configuration obtained after applying the relax-tfMC/MC protocol in Figure~\ref{fig:Figure06}(c).

We denote by $x$ the global Ag fraction that sets the coherent reference lattice parameters $(a_0,c_0)$, 
and by $y$ the local Ag to Cu composition that is explored by the Monte Carlo protocol.
Following Larche--Cahn elasticity \citep{Larche1973}, a local composition deviation $(y-x)$ generates a strain
and within linear elasticity, the corresponding elastic energy density is
\begin{equation}
E_{\rm el}(x,y)=\frac{1}{2}(y-x)^2\,\eta^T C \eta,
\end{equation}
with $C$ the stiffness tensor (Figure~\ref{fig:Figure04}(a)). The chemical expansion tensor is defined as
\begin{equation}
\eta_{ij}=\left.\frac{\partial \varepsilon_{ij}}{\partial y}\right|_{\sigma=0,T},
\end{equation}
and is obtained from the stress-free lattice parameters $a(y)$ and $c(y)$ via $\varepsilon_{xx}=(a-a_0)/a_0$ and $\varepsilon_{zz}=(c-c_0)/c_0$. Thus, $\eta$ quantitatively links local composition fluctuations to coherency strain and its elastic energy.

As an alternative to the introduced semi-analytical model, we compute the elastic contribution fully 
numerically from ML energies at fixed composition. For each $y$ we evaluate (i) the coherent energy 
$E_{\rm coh}^x(y)$ by fixing the cell to the coherent reference $(a_0,c_0)$ of the 
global composition $x$ and relaxing atomic positions only, and (ii) the fully relaxed 
reference energy $E_{\rm rel}(y)$ by relaxing both cell and atomic positions. The coherency elastic energy is then
\begin{equation}
\Delta H_{\rm el}(x,y)=E_{\rm coh}^{(x)}(y)-E_{\rm rel}(y),
\end{equation}
which satisfies $E_{\rm el}(x,x)=0$ by construction.

In Figures~\ref{fig:Figure07}\,(b,e) we present the mixing enthalpy $\Delta H_{\mathrm{mix}}$, calculated without $E_{\mathrm{el}}$, the elastic mixing enthalpy $\Delta H_{\mathrm{el}}$ for different choices of coherent reference lattice parameters, and the total enthalpy $\Delta H_{\mathrm{total}}=\Delta H_{\mathrm{mix}}+\Delta H_{\mathrm{el}}$, as a function of the Ag-Cu ratio. In Figure~\ref{fig:Figure07}(b) we consider the reference lattice parameters of the fully mixed (Ag$_{0.5}$Cu$_{0.5}$)GaSe$_2$. We find that $ \Delta H_{\rm el}$ is negative and that its magnitude exceeds $ \Delta H_{\mathrm{mix}}$ over a wide compositional range, so that $\Delta H_{\rm total}$ becomes negative and is dominated by the elastic driving force. Moreover, we compute $\Delta H_{\rm el}$  from both (i) the semi-analytical linear-elastic expression and (ii) the fully numerical coherent--relaxed energy difference, and  find excellent agreement. In the remainder, all computations are performed using the fully numerical approach, while the semi-analytical model is used to discuss the underlying physics.

Next, to investigate the incoherent-interface scenario in Figure~\ref{fig:Figure06}(c), we evaluate $H_{\rm el}$ using the two pure reference lattice parameters AgGaSe$_2$ and CuGaSe$_2$. Since we aim to quantify the driving force from a mixed to a demixed configuration, we take the homogeneous $0.5$ composition as the starting point. This starting point is indicated by the arrows in Figure~\ref{fig:Figure07}(d), whose directions correspond to the composition changes that reduce the coherency strain in each region. In this case $\Delta H_{\rm el}$ is positive and exceeding $\Delta H_{\rm mix}$ at $0.5$. Consequently, $\Delta H_{\rm total}$ increases further, indicating an enhanced driving force consistent with the phase separation observed for the incoherent interface after applying the relax-tfMC/MC protocol, see Figure~\ref{fig:Figure06}(c).

From Figures~\ref{fig:Figure07}(a,d) we observe that the stress-free local composition depends on the imposed coherent reference lattice parameters. We therefore modify the Gibbs free energy as
\begin{equation}
   \Delta G(x,y)_{\mathrm{mix}}= \Delta H_{\mathrm{mix}}(y)+ \Delta H_{\mathrm{el}}(x,y)-T \Delta S_{\mathrm{mix}}(y),
\end{equation}
where $x$ denotes the reference composition that defines the coherent lattice parameters $(a_0,c_0)$, and $y$ denotes the local Ag/Cu composition.

We first consider the case where $(a_0,c_0)$ corresponds to the homogeneous (Ag$_{0.5}$Cu$_{0.5}$)GaSe$_2$ reference, see Figure~\ref{fig:Figure07}(c). In this case, the Gibbs free energy of mixing computed from Eq.~8 is negative and exhibits no secondary local minima, implying the absence of a miscibility gap. This prediction, which explicitly accounts for the elastic energy associated with the mismatch between the local composition and the global coherent lattice parameters, differs qualitatively from phase-stability assessments based on bulk {\it ab initio} calculations \citep{Chen2007,Sopiha2020}. In contrast, when the end-point reference lattice parameters (AgGaSe$_2$ and CuGaSe$_2$) are imposed and the overall initial composition is the homogeneous $0.5$, a miscibility gap emerges. These results are consistent with the incoherent-interface scenario shown in Figure~\ref{fig:Figure06}(c).

These results contribute to the explanation why co-evaporated Ag$_x$Cu$_{1-x}$(In,Ga)Se$_2$ devices can exhibit stable performance: 
coherent strain energetically favors mixing under the thin-film coherency constraint. Apparent contradictions with bulk-based 
assessments may arise from fabrication processing routes that relax coherency constrains (e.g., via dislocations, grain boundaries, etc.), 
thereby favoring pure phase coherent lattices.

\section{Conclusions and future work\label{sec:VI}}

In this work, we develop a ML potential to explore the Ag-Cu compositional space and identify stable Ag$_x$Cu$_{1-x}$GaSe$_2$ configurations. 
The ML potential reproduces structural and elastic properties in good agreement with available {\it ab initio} and experimental data and 
enables the calculation of thermodynamic observables and the phase diagram, consistent with prior computational studies (Refs. \citep{Chen2007,Sopiha2020}).

Atomistic Monte Carlo simulations further show that coherency constraints couple chemistry with mechanical properties: a mismatch between local 
chemical composition and coherent lattice constants generates a substantial elastic energy that can dominate the mixing thermodynamics. Accounting for 
this coherency-strain contribution leads to complete Ag-Cu miscibility under coherent boundary conditions.

These findings explain the apparently conflicting experimental reports on the Ag-Cu miscibility: the observed behavior depends on 
processing-induced mechanical boundary conditions. Co-evaporated thin films can sustain a largely coherent lattice over relevant length scales, 
favoring solid-solution formation during thermal processing, whereas ingot-based routes may more readily relax mismatch (e.g., via defects and 
local lattice accommodation), promoting demixing.

Generally, this physical mechanism should be relevant to chalcopyrite alloys with small chemical driving forces for mixing. Moreover, 
alkali substitutions (Na, K) on Ag-Cu sites exhibit mixing enthalpies comparable to those in Figure \ref{fig:Figure05}(a) \citep{Aboulfadl2021}, 
suggesting that phase stability predictions can change qualitatively when elastic contributions are neglected. 

\section*{CRediT authorship contribution statement}

Vasilios Karanikolas: Writing – review and editing, 
Writing – original draft, Resources, Visualization, Validation, Software, Methodology, Investigation, Formal analysis, Conceptualization. 

Delwin Perera: Writing – review and editing, Visualization, Validation, Software, Methodology, Investigation, Formal analysis, Conceptualization. 

Linus Erhard: Visualization, Software, Methodology. 

Jochen Rohrer: Writing – review and editing, Validation, Software, Methodology, Investigation, Formal analysis. 

Karsten Albe: Writing – review and editing, Validation, Supervision, Resources, Project administration, Methodology, Funding acquisition, Formal analysis, Data curation, Conceptualization.

\section*{Declaration of competing interest}
The authors declare that they have no known competing financial interests or personal relationships
that could have appeared to influence the work reported in this paper.

\section*{Acknowledgements}

VK gratefully acknowledge the computing time provided to them at the NHR Center NHR4CES at TU Darmstadt (project number p0022202). This is funded by the Federal Ministry of Research, Technology and Space, and the state governments participating on the basis of the resolutions of the GWK for national high performance computing at universities. The authors gratefully acknowledge the computing time provided on the high-performance computer HoreKa by the National High-Performance Computing Center at KIT (NHR@KIT). This center is jointly supported by the Federal Ministry of Education and Research and the Ministry of Science, Research and the Arts of Baden-Württemberg, as part of the National High-Performance Computing (NHR) joint funding program. HoreKa is partly funded by the German Research Foundation (DFG).

\section*{Funding}
VK and KA gratefully acknowledge funding by the NHR4CES research project as part of SDL Materials Design. DP and KA research is funded by the Deutsche Forschungsgemeinschaft (DFG, German Research Foundation) – Project number 414750661.

\section*{Data availability}

The research data supporting this publication are openly available in Zenodo at \url{https://doi.org/10.5281/zenodo.19183056}.

\section*{Declaration of generative AI and AI-assisted technologies in
the manuscript preparation process}
During the preparation of this work, the authors used ChatGPT-5.2 (OpenAI) for code review and editorial control. After using this tool/service, the authors reviewed and edited the content as needed and take full responsibility for the content of the published article.


\begin{thebibliography}{99}

\bibitem{Oliveira2022}
A.J.N. Oliveira, J.P. Teixeira, D. Ramos, P.A. Fernandes, P.M.P. Salom\'e, Exploiting the Optical Limits of Thin-Film Solar Cells: A Review on Light Management Strategies in Cu(In,Ga)Se2,  Adv. Photonics Res. 3 (2022) 2100190.  \href{https://doi.org/10.1002/adpr.202100190}{https://doi.org/10.1002/adpr.202100190}.


\bibitem{Mufti2020}
N. Mufti, T. Amrillah, A. Taufiq, Sunaryono, Aripriharta, M. Diantoro, Zulhadjri, H. Nur, Review of CIGS-based solar cells manufacturing by structural engineering, Sol. Energy 207 (2020) 1146--1157. \href{https://doi.org/10.1016/j.solener.2020.07.065}{https://doi.org/10.1016/j.solener.2020.07.065}.

\bibitem{Nakamura2019}
M. Nakamura, K. Yamaguchi, Y. Kimoto, Y. Yasaki, T. Kato, H. Sugimoto, Cd-Free Cu(In,Ga)(Se,S)$_2$ Thin-Film Solar Cell With Record Efficiency of 23.35\%, IEEE J. Photovolt. 9 (2019) 1863--1867. \href{https://doi.org/10.1109/JPHOTOV.2019.2937218}{https://doi.org/10.1109/JPHOTOV.2019.2937218}.

\bibitem{Keller2024}
J. Keller, K. Kiselman, O. Donzel-Gargand, N.M. Martin, M. Babucci, O. Lundberg, E. Wallin, L. Stolt, M. Edoff, High-concentration silver alloying and steep back-contact gallium grading enabling copper indium gallium selenide solar cell with 23.6\% efficiency, Nat. Energy 9 (2024) 467--478. \href{https://doi.org/10.1038/s41560-024-01472-3}{https://doi.org/10.1038/s41560-024-01472-3}.

\bibitem{Erslev2011}
P.T. Erslev, J. Lee, G.M. Hanket, W.N. Shafarman, J.D. Cohen, The electronic structure of Cu(In$_{1-x}$Ga$_x$)Se$_2$ alloyed with silver, Thin Solid Films 519 (2011) 7296--7299. \href{https://doi.org/10.1016/j.tsf.2011.01.368}{https://doi.org/10.1016/j.tsf.2011.01.368}.

\bibitem{Yang2021}
S.-C. Yang, J. Sastre, M. Krause, X. Sun, R. Hertwig, M. Ochoa, A.N. Tiwari, R. Carron, Silver-Promoted High-Performance (Ag,Cu)(In,Ga)Se$_2$ Thin-Film Solar Cells Grown at Very Low Temperature, Sol. RRL 5 (2021) 2100108. \href{https://doi.org/10.1002/solr.202100108}{https://doi.org/10.1002/solr.202100108}.

\bibitem{Abou-Ras2016}
D. Abou-Ras, N. Schafer, T. Rissom, M.N. Kelly, J. Haarstrich, C. Ronning, G.S. Rohrer, A.D. Rollett, Grain-boundary character distribution and correlations with electrical and optoelectronic properties of CuInSe$_2$ thin films, Acta Mater. 118 (2016) 244--252. \href{https://doi.org/10.1016/j.actamat.2016.07.042}{https://doi.org/10.1016/j.actamat.2016.07.042}.

\bibitem{Valdes2019}
N. Valdes, J.W. Lee, W. Shafarman, Comparison of Ag and Ga alloying in low bandgap CuInSe$_2$-based solar cells, Sol. Energy Mater. Sol. Cells 195 (2019) 155--159. \href{https://doi.org/10.1016/j.solmat.2019.02.022}{https://doi.org/10.1016/j.solmat.2019.02.022}.

\bibitem{Kim2016}
K. Kim, J.W. Park, J.S. Yoo, J. Sik Cho, H.-D. Lee, J.H. Yun, Ag incorporation in low-temperature grown Cu(In,Ga)Se$_2$ solar cells using Ag precursor layers, Sol. Energy Mater. Sol. Cells 146 (2016) 114--120. \href{https://doi.org/10.1016/j.solmat.2015.11.028}{https://doi.org/10.1016/j.solmat.2015.11.028}.


\bibitem{Gu2024}
Y. Gu, C. Zhou, W. Chen, Y. Zhang, Y. Yao, Z. Zhou, Y. Sun, W. Liu, Silver (Ag) substitution in Cu(In,Ga)Se$_2$ solar cell: insights into current challenges and future prospects, Appl. Phys. A 130 (2024) 636. \href{https://doi.org/10.1007/s00339-024-07796-x}{https://doi.org/10.1007/s00339-024-07796-x}.

\bibitem{Kim2018}
K. Kim, S.K. Ahn, J.H. Choi, J. Yoo, Y.J. Eo, J.S. Cho, A. Cho, J. Gwak, S. Song, D.H. Cho, Y.D. Chung, J.H. Yun, Highly Efficient Ag-Alloyed Cu(In,Ga)Se$_2$ Solar Cells with Wide Bandgaps and Their Application to Chalcopyrite-Based Tandem Solar Cells, Nano Energy 48 (2018) 345--352. \href{https://doi.org/10.1016/j.nanoen.2018.03.052}{https://doi.org/10.1016/j.nanoen.2018.03.052}.

\bibitem{Zhang2020}
Y. Zhang, Z. Hu, S. Lin, S. Cheng, Z. He, C. Wang, Z. Zhou, Y. Sun, W. Liu, Facile Silver-Incorporated Method of Tuning the Back Gradient of Cu(In,Ga)Se$_2$ Films, ACS Appl. Energy Mater. 3 (2020) 9963--9971. \href{https://doi.org/10.1021/acsaem.0c01644}{https://doi.org/10.1021/acsaem.0c01644}.

\bibitem{Krause2023}
M. Krause, S.C. Yang, S. Moser, S. Nishiwaki, A.N. Tiwari, R. Carron, Silver-Alloyed Low-Bandgap CuInSe$_2$ Solar Cells for Tandem Applications, Sol. RRL 7 (2023) 2201122. \href{https://doi.org/10.1002/solr.202201122}{https://doi.org/10.1002/solr.202201122}.

\bibitem{Siebentritt2023}
S. Siebentritt, T.P. Weiss, Chalcopyrite solar cells---state-of-the-art and options for improvement, Sci. China Phys. Mech. Astron. 66 (2023) 217301. \href{https://doi.org/10.1007/s11433-022-2001-4}{https://doi.org/10.1007/s11433-022-2001-4}.

\bibitem{Chen2007}
S. Chen, X.G. Gong, S.-H. Wei, Band-structure anomalies of the chalcopyrite semiconductors CuGaX$_2$ versus AgGaX$_2$ (X = S and Se) and their alloys, Phys. Rev. B 75 (2007) 205209. \href{https://doi.org/10.1103/PhysRevB.75.205209}{https://doi.org/10.1103/PhysRevB.75.205209}.

\bibitem{Sopiha2020}
K.V. Sopiha, J.K. Larsen, O. Donzel-Gargand, F. Khavari, J. Keller, M. Edoff, C. Platzer-Bj\"orkman, C. Persson, J.J.S. Scragg, Thermodynamic stability, phase separation and Ag grading in (Ag,Cu)(In,Ga)Se$_2$ solar absorbers, J. Mater. Chem. A 8 (2020) 8740--8751. \href{https://doi.org/10.1039/D0TA00363H}{https://doi.org/10.1039/D0TA00363H}.

\bibitem{Robbins1973}
M. Robbins, J.C. Phillips, V.G. Lambrecht Jr., Solid solution formation in the systems CuM$^{\mathrm{III}}$X$_2$--AgM$^{\mathrm{III}}$X$_2$ where M$^{\mathrm{III}}$ = Al, Ga, In and X = S, Se, J. Phys. Chem. Solids 34 (1973) 1205--1209.
\href{https://doi.org/10.1016/S0022-3697(73)80210-0}{https://doi.org/10.1016/S0022-3697(73)80210-0}.

\bibitem{Avon1984}
J.E. Avon, K. Yoodee, J.C. Woolley, Solid solution, lattice parameter values, and effects of composition in Ag$_x$Cu$_{1-x}$InSe$_2$ chalcopyrite, J. Appl. Phys. 55 (1984) 524--535.  \href{https://doi.org/10.1063/1.333058}{https://doi.org/10.1063/1.333058}.

\bibitem{Albornoz2014}
J.G. Albornoz, R.M. Rojas L., J.M. Merino, M. Leon, Structural, thermal and electrical properties of the semiconductor system Ag$_{1-x}$Cu$_x$InSe$_2$, J. Phys. Chem. Solids 75 (2014) 1--7. \href{https://doi.org/10.1016/j.jpcs.2013.08.003}{https://doi.org/10.1016/j.jpcs.2013.08.003}.

\bibitem{Hanket2009}
G.M. Hanket, J.H. Boyle, W.N. Shafarman, Characterization and Device Performance of (AgCu)(InGa)Se$_2$ Absorber Layers, in: Proc. 34th IEEE Photovoltaic Specialists Conf. (PVSC), Philadelphia, PA, USA, 2009, pp. 001240--001245. \href{https://doi.org/10.1109/PVSC.2009.5411241}{https://doi.org/10.1109/PVSC.2009.5411241}.

\bibitem{Hanket2010}
G.M. Hanket, J.H. Boyle, W.N. Shafarman, G. Teeter, Wide-bandgap (AgCu)(InGa)Se$_2$ absorber layers deposited by three-stage co-evaporation, in: Proc. 35th IEEE Photovoltaic Specialists Conf. (PVSC), Honolulu, HI, USA, 2010, pp. 003425--003429. \href{https://doi.org/10.1109/PVSC.2010.5614576}{https://doi.org/10.1109/PVSC.2010.5614576}.

\bibitem{Boyle2011}
J.H. Boyle, B.E. McCandless, G.M. Hanket, W.N. Shafarman, Structural characterization of the (AgCu)(InGa)Se$_2$ thin film alloy system by x-ray diffraction, Thin Solid Films 519 (2011) 7292--7295. \href{https://doi.org/10.1016/j.tsf.2011.01.138}{https://doi.org/10.1016/j.tsf.2011.01.138}.

\bibitem{Chen2014}
L. Chen, J.W. Lee, W.N. Shafarman, The Comparison of (Ag,Cu)(In,Ga)Se$_2$ and Cu(In,Ga)Se$_2$ Thin Films Deposited by Three-Stage Coevaporation, IEEE J. Photovolt. 4 (2014) 447--451. \href{https://doi.org/10.1109/JPHOTOV.2013.2280471}{https://doi.org/10.1109/JPHOTOV.2013.2280471}.

\bibitem{Boyle2014}
J.H. Boyle, B.E. McCandless, W.N. Shafarman, R.W. Birkmire, Structural and optical properties of (Ag,Cu)(In,Ga)Se$_2$ polycrystalline thin film alloys, J. Appl. Phys. 115 (2014) 223504. \href{https://doi.org/10.1063/1.4880243}{https://doi.org/10.1063/1.4880243}.

\bibitem{Erhard2022}
L.C. Erhard, J. Rohrer, K. Albe, V.L. Deringer, A machine-learned interatomic potential for silica and its relation to empirical models, npj Comput. Mater. 8 (2022) 90. \href{https://doi.org/10.1038/s41524-022-00768-w}{https://doi.org/10.1038/s41524-022-00768-w}.

\bibitem{Mortazavi2023}
B. Mortazavi, X. Zhuang, T. Rabczuk, A.V. Shapeev, Atomistic modeling of the mechanical properties: the rise of machine learning interatomic potentials, Mater. Horiz. 10 (2023) 1956--1968. \href{https://doi.org/10.1039/D3MH00125C}{https://doi.org/10.1039/D3MH00125C}.

\bibitem{Erhard2024}
L.C. Erhard, J. Rohrer, K. Albe, V.L. Deringer, Modelling atomic and nanoscale structure in the silicon--oxygen system through active machine learning, Nat. Commun. 15 (2024) 1927. \href{https://doi.org/10.1038/s41467-024-45840-9}{https://doi.org/10.1038/s41467-024-45840-9}.

\bibitem{Jacobs2025}
R. Jacobs, D. Morgan, S. Attarian, J. Meng, C. Shen, Z. Wu, C.Y. Xie, J.H. Yang, N. Artrith, B. Blaiszik, G. Ceder, K. Choudhary, G. Csanyi, E.D. Cubuk, B. Deng, R. Drautz, X. Fu, J. Godwin, V. Honavar, O. Isayev, A. Johansson, B. Kozinsky, S. Martiniani, S.P. Ong, I. Poltavsky, K. Schmidt, S. Takamoto, A.P. Thompson, J. Westermayr, B.M. Wood, A practical guide to machine learning interatomic potentials -- Status and future, Curr. Opin. Solid State Mater. Sci. 35 (2025) 101214. \href{https://doi.org/10.1016/j.cossms.2025.101214}{https://doi.org/10.1016/j.cossms.2025.101214}.

\bibitem{Baskaran2015}
A. Baskaran, C. Ratsch, P. Smereka, Modeling the elastic energy of alloys: Potential pitfalls of Fourier-based multilattice methods, Phys. Rev. E 92 (2015) 062406. \href{https://doi.org/10.1103/PhysRevE.92.062406}{https://doi.org/10.1103/PhysRevE.92.062406}.

\bibitem{Dereli1992}
G. Dereli, Stillinger-Weber type potentials in Monte Carlo simulations of amorphous silicon, Mol. Simul. 8 (1992) 351-360. \href{https://doi.org/10.1080/08927029208022490}{https://doi.org/10.1080/08927029208022490}.

\bibitem{Timonova2010}
M. Timonova, J. Groenewegen, B.J. Thijsse, Modeling diffusion and phase transitions by a uniform-acceptance force-bias Monte Carlo method, Phys. Rev. B 81 (2010) 144107. \href{https://doi.org/10.1103/PhysRevB.81.144107}{https://doi.org/10.1103/PhysRevB.81.144107}.

\bibitem{Mees2012}
M.J. Mees, G. Pourtois, E.C. Neyts, B.J. Thijsse, A. Stesmans, Uniform-acceptance force-bias Monte Carlo method with time scale to study solid-state diffusion, Phys. Rev. B 85 (2012) 134301. \href{https://doi.org/10.1103/PhysRevB.85.134301}{https://doi.org/10.1103/PhysRevB.85.134301}.

\bibitem{Bal2014}
K.M. Bal, E.C. Neyts, On the time scale associated with Monte Carlo simulations, J. Chem. Phys. 141 (2014) 204104. \href{https://doi.org/10.1063/1.4902136}{https://doi.org/10.1063/1.4902136}.

\bibitem{Eshelby1957}
J.D. Eshelby, The determination of the elastic field of an ellipsoidal inclusion, and related problems, Proc. R. Soc. Lond. A 241 (1957) 376--396. \href{https://doi.org/10.1098/rspa.1957.0133}{https://doi.org/10.1098/rspa.1957.0133}.

\bibitem{Larche1973}
F. Larche, J.W. Cahn, A linear theory of thermochemical equilibrium of solids under stress, Acta Metall. 21 (1973) 1051--1063. \href{https://doi.org/10.1016/0001-6160(73)90021-7}{https://doi.org/10.1016/0001-6160(73)90021-7}.

\bibitem{Larche1982}
F.C. Larche, J.W. Cahn, The effect of self-stress on diffusion in solids, Acta Metall. 30 (1982) 1835--1845. \href{https://doi.org/10.1016/0001-6160(82)90023-2}{https://doi.org/10.1016/0001-6160(82)90023-2}.

\bibitem{Larche1984}
F.C. Larche, J.W. Cahn, The Interactions of Composition and Stress in Crystalline Solids, J. Res. Natl. Bur. Stand. 89 (1984) 467--500. \href{https://doi.org/10.6028/jres.089.026}{https://doi.org/10.6028/jres.089.026}.

\bibitem{Fratzl1999}
P. Fratzl, O. Penrose, J.L. Lebowitz, Modeling of phase separation in alloys with coherent elastic misfit, J. Stat. Phys. 95 (1999) 1429--1503. \href{https://doi.org/10.1023/A:1004587425006}{https://doi.org/10.1023/A:1004587425006}.

\bibitem{Kamachali2022}
R.D. Kamachali, L. Wang, Elastic energy of multi-component solid solutions and strain origins of phase stability in high-entropy alloys, Scr. Mater. 206 (2022) 114226. \href{https://doi.org/10.1016/j.scriptamat.2021.114226}{https://doi.org/10.1016/j.scriptamat.2021.114226}.

\bibitem{Weissmuller2022}
J. Weissmuller, S. Shi, Giant compliance and spontaneous buckling of beams containing mobile solute atoms, Acta Mater. 227 (2022) 117696. \href{https://doi.org/10.1016/j.actamat.2022.117696}{https://doi.org/10.1016/j.actamat.2022.117696}.

\bibitem{Robbins1975}
M. Robbins, V.G. Lambrecht Jr., Solid solution formation in chalcopyrite systems of the type AgInX$_2$--AgGaX$_2$, X = S, Se, Te, J. Solid State Chem. 15 (1975) 167--170.  \href{https://doi.org/10.1016/0022-4596(75)90240-6}{https://doi.org/10.1016/0022-4596(75)90240-6}.

\bibitem{Drautz2019}
R. Drautz, Atomic cluster expansion for accurate and transferable interatomic potentials, Phys. Rev. B 99 (2019) 014104. \href{https://doi.org/10.1103/PhysRevB.99.014104}{https://doi.org/10.1103/PhysRevB.99.014104}.

\bibitem{Lysogorskiy2021}
Y. Lysogorskiy, C. van der Oord, A. Bochkarev, S. Menon, M. Rinaldi, T. Hammerschmidt, M. Mrovec, A.P. Thompson, G. Csanyi, C. Ortner, R. Drautz, Performant implementation of the atomic cluster expansion (PACE) and application to copper and silicon, npj Comput. Mater. 7 (2021) 97.  \href{https://doi.org/10.1038/s41524-021-00559-9}{https://doi.org/10.1038/s41524-021-00559-9}.

\bibitem{Leimeroth2025}
N. Leimeroth, L.C. Erhard, K. Albe, J. Rohrer, Machine-learning interatomic potentials from a users perspective: a comparison of accuracy, speed and data efficiency, Modell. Simul. Mater. Sci. Eng. 33 (2025) 065012. \href{https://doi.org/10.1088/1361-651X/adf56d}{https://doi.org/10.1088/1361-651X/adf56d}.

\bibitem{Qamar2023}
M. Qamar, M. Mrovec, Y. Lysogorskiy, A. Bochkarev, R. Drautz, Atomic cluster expansion for quantum-accurate large-scale simulations of carbon, J. Chem. Theory Comput. 19 (2023) 5151--5167. \href{https://doi.org/10.1021/acs.jctc.2c01149}{https://doi.org/10.1021/acs.jctc.2c01149}.

\bibitem{Podryabinkin2017}
E.V. Podryabinkin, A.V. Shapeev, Active learning of linearly parametrized interatomic potentials, Comput. Mater. Sci. 140 (2017) 171--180. \href{https://doi.org/10.1016/j.commatsci.2017.08.031}{https://doi.org/10.1016/j.commatsci.2017.08.031}. 

\bibitem{Novikov2021}
I.S. Novikov, K. Gubaev, E.V. Podryabinkin, A.V. Shapeev, The MLIP package: moment tensor potentials with MPI and active learning, Mach. Learn.: Sci. Technol. 2 (2021) 025002. \href{https://doi.org/10.1088/2632-2153/abc9fe}{https://doi.org/10.1088/2632-2153/abc9fe}.

\bibitem{Lysogorskiy2023}
Y. Lysogorskiy, A. Bochkarev, M. Mrovec, R. Drautz, Active learning strategies for atomic cluster expansion models, Phys. Rev. Mater. 7 (2023) 043801. \href{https://doi.org/10.1103/PhysRevMaterials.7.043801}{https://doi.org/10.1103/PhysRevMaterials.7.043801}.

\bibitem{Bochkarev2022}
A. Bochkarev, Y. Lysogorskiy, S. Menon, M. Qamar, M. Mrovec, R. Drautz, Efficient parametrization of the atomic cluster expansion, Phys. Rev. Mater. 6 (2022) 013804. \href{https://doi.org/10.1103/PhysRevMaterials.6.013804}{https://doi.org/10.1103/PhysRevMaterials.6.013804}.

\bibitem{Kresse1994}
G. Kresse, J. Furthm\"uller, J. Hafner, Theory of the crystal structures of selenium and tellurium, Phys. Rev. B 50 (1994) 13181--13185. \href{https://doi.org/10.1103/PhysRevB.50.13181}{https://doi.org/10.1103/PhysRevB.50.13181}.

\bibitem{Kresse1996}
G. Kresse, J. Furthm\"uller, Efficiency of ab-initio total energy calculations for metals and semiconductors using a plane-wave basis set, Comput. Mater. Sci. 6 (1996) 15--50. \href{https://doi.org/10.1016/0927-0256(96)00008-0}{https://doi.org/10.1016/0927-0256(96)00008-0}.

\bibitem{Perdew1996}
J.P. Perdew, K. Burke, M. Ernzerhof, Generalized Gradient Approximation Made Simple, Phys. Rev. Lett. 77 (1996) 3865--3868. \href{https://doi.org/10.1103/PhysRevLett.77.3865}{https://doi.org/10.1103/PhysRevLett.77.3865}.

\bibitem{Kresse1999}
G. Kresse, D. Joubert, From ultrasoft pseudopotentials to the projector augmented-wave method, Phys. Rev. B 59 (1999) 1758--1775. \href{https://doi.org/10.1103/PhysRevB.59.1758}{https://doi.org/10.1103/PhysRevB.59.1758}.

\bibitem{Thompson2022}
A.P. Thompson, H.M. Aktulga, R. Berger, D.S. Bolintineanu, W.M. Brown, P.S. Crozier, P.J. in 't Veld, A. Kohlmeyer, S.G. Moore, T.D. Nguyen, R. Shan, M.J. Stevens, J. Tranchida, C. Trott, S.J. Plimpton, LAMMPS -- a flexible simulation tool for particle-based materials modeling at the atomic, meso, and continuum scales, Comp. Phys. Comm. 271 (2022) 108171. \href{https://doi.org/10.1016/j.cpc.2021.108171}{https://doi.org/10.1016/j.cpc.2021.108171}.

\bibitem{Sadigh2012}
B. Sadigh, P. Erhart, A. Stukowski, A. Caro, E. Martinez, L. Zepeda-Ruiz, Scalable parallel Monte Carlo algorithm for atomistic simulations of precipitation in alloys, Phys. Rev. B 85 (2012) 184203. \href{https://doi.org/10.1103/PhysRevB.85.184203}{https://doi.org/10.1103/PhysRevB.85.184203}.

\bibitem{Neyts2013}
E.C. Neyts, A. Bogaerts, Combining molecular dynamics with Monte Carlo simulations: implementations and applications, Theor. Chem. Acc. 132 (2013) 1320. \href{https://doi.org/10.1007/s00214-012-1320-x}{https://doi.org/10.1007/s00214-012-1320-x}

\bibitem{SupplementalMaterials}
In the Supplemental Materials we provide validation plots for the ML potential, energy-volume curves for different Ag$_x$Cu$_{1-x}$GaSe$_2$ phases, the temperature dependence of the lattice parameters, the anion displacement, present the relax-tfMC/MD protocol and its application to a spherical inclusion structure. \href{https://doi.org/10.1007/xxx}{https://doi.org/xxx}.

\bibitem{Grimsditch1975}
M.H. Grimsditch, G.D. Holah, Brillouin scattering and elastic moduli of silver thiogallate (AgGaS$_2$), Phys. Rev. B 12 (1975) 4377. \href{https://doi.org/10.1103/PhysRevB.12.4377}{https://doi.org/10.1103/PhysRevB.12.4377}.

\bibitem{Neumann1986}
H. Neumann, Bulk modulus-volume relationship in ternary chalcopyrite compounds, Phys. Status Solidi A 96 (1986) K121--K125. \href{https://doi.org/10.1002/pssa.2210960245}{https://doi.org/10.1002/pssa.2210960245}. 

\bibitem{Neumann1983}
H. Neumann, Simple theoretical estimate of surface energy, bulk modulus, and atomization energy of A$^{\mathrm{I}}$B$^{\mathrm{III}}$C$_2^{\mathrm{VI}}$ compounds, Cryst. Res. Technol. 18 (1983) 665--670. \href{https://doi.org/10.1002/crat.2170180517}{https://doi.org/10.1002/crat.2170180517}.

\bibitem{Fernandez1990}
B. Fernandez, S.M. Wasim, Sound Velocities and Elastic Moduli in CuInTe$_2$ and CuInSe$_2$, Phys. Status Solidi A 122 (1990) 235.  \href{https://doi.org/10.1002/pssa.2211220122}{https://doi.org/10.1002/pssa.2211220122}.

\bibitem{Garbato1987}
L. Garbato, F. Ledda, R. Rucci, Structural distortions and polymorphic behaviour in ABC$_2$ chalcopyrites, Prog. Cryst. Growth Charact. 15 (1987) 1--41.   \href{https://doi.org/10.1016/0146-3535(87)90008-6}{https://doi.org/10.1016/0146-3535(87)90008-6}.

\bibitem{Falk2023}
H.H. Falk, S. Eckner, M. Seifert, K. Ritter, S. Levcenko, T. Pfeiffelmann, E. Welter, S. Botti, W.N. Shafarman, C.S. Schnohr, Peculiar bond length dependence in (Ag,Cu)GaSe$_2$ alloys and its impact on the bandgap bowing, APL Mater. 11 (2023) 111105. \href{https://doi.org/10.1063/5.0164407}{https://doi.org/10.1063/5.0164407}.

\bibitem{Shay1975}
J.L. Shay, J.H. Wernick, Ternary Chalcopyrite Semiconductors: Growth, Electronic Properties and Applications, Pergamon Press, Oxford, 1975.

\bibitem{Verma2012}
A.S. Verma, S. Sharma, R. Bhandari, B.K. Sarkar, V.K. Jindal, Elastic properties of chalcopyrite structured solids, Mater. Chem. Phys. 132 (2012) 416--420. \href{https://doi.org/10.1016/j.matchemphys.2011.11.047}{https://doi.org/10.1016/j.matchemphys.2011.11.047}.

\bibitem{Kushwaha2019}
A.K. Kushwaha, C.-G. Ma, M.G. Brik, S. Bin Omran, R. Khenata, Zone-center phonons and elastic properties of ternary chalcopyrite ABSe$_2$ (A = Cu and Ag; B = Al, Ga and In), Mater. Chem. Phys. 227 (2019) 324--331. \href{https://doi.org/10.1016/j.matchemphys.2019.02.024}{https://doi.org/10.1016/j.matchemphys.2019.02.024}.

\bibitem{Stukowski2010}
A. Stukowski, Visualization and analysis of atomistic simulation data with OVITO---the Open Visualization Tool, Modell. Simul. Mater. Sci. Eng. 18 (2010) 015012. \href{https://doi.org/10.1088/0965-0393/18/1/015012}{https://doi.org/10.1088/0965-0393/18/1/015012}.

\bibitem{Aboulfadl2021}
H. Aboulfadl, K.V. Sopiha, J. Keller, J.K. Larsen, J.J.S. Scragg, C. Persson, M. Thuvander, M. Edoff, Alkali Dispersion in (Ag,Cu)(In,Ga)Se$_2$ Thin Film Solar Cells---Insight from Theory and Experiment, ACS Appl. Mater. Interfaces 13 (2021) 7188--7199. \href{https://doi.org/10.1021/acsami.0c20539}{https://doi.org/10.1021/acsami.0c20539}.


\end{thebibliography}
\end{document}